\documentclass[letterpaper]{article} % DO NOT CHANGE THIS
\usepackage{aaai24}  % DO NOT CHANGE THIS
\usepackage{times}  % DO NOT CHANGE THIS
\usepackage{helvet}  % DO NOT CHANGE THIS
\usepackage{courier}  % DO NOT CHANGE THIS
\usepackage[hyphens]{url}  % DO NOT CHANGE THIS
\usepackage{graphicx} % DO NOT CHANGE THIS
\usepackage{natbib}  % DO NOT CHANGE THIS AND DO NOT ADD ANY OPTIONS TO IT
\usepackage{caption} % DO NOT CHANGE THIS AND DO NOT ADD ANY OPTIONS TO IT
\usepackage{algorithm}
\usepackage{algorithmic}
\usepackage{newfloat}
\usepackage{listings}
\DeclareCaptionStyle{ruled}{labelfont=normalfont,labelsep=colon,strut=off} % DO NOT CHANGE THIS
\floatstyle{ruled}
\newfloat{listing}{tb}{lst}{}
\floatname{listing}{Listing}
\usepackage{url}            % simple URL typesetting
\usepackage{booktabs}       % professional-quality tables
\usepackage{amsfonts}       % blackboard math symbols
\usepackage{nicefrac}       % compact symbols for 1/2, etc.
\usepackage{microtype}      % microtypography
\usepackage{amsfonts}
\usepackage{graphicx}
\usepackage{mathrsfs}
\usepackage{amsmath}
\usepackage{amssymb}
\usepackage{float,color,natbib}
\usepackage{enumerate}
\usepackage{tabularx}
\usepackage{lineno}
\usepackage{xr}
\usepackage{diagbox}
\usepackage{booktabs}
\usepackage{algorithm}

\usepackage{algorithmic}
\def\squarebox#1{\hbox to #1{\hfill\vbox to #1{\vfill}}}
\def\boxit#1{\vbox{\hrule\hbox{\vrule\kern6pt
          \vbox{\kern6pt#1\kern6pt}\kern6pt\vrule}\hrule}}

\def\E{\mathbb {E}}
\def\Th{\widehat{\mathcal{T}}}

\newtheorem{assu}{Assumption}

\newtheorem{thm}{Theorem}
\newtheorem{pro}{Proposition}
\newtheorem{lemma}{Lemma}
\newtheorem{coro}{Corollary}

\def\remark#1{\noindent{\bf Remark #1\ }}

\newcommand{\bm}{\boldsymbol}

\def\X{{\bf X}}

\def\th{\widehat{\tau}}

\newcommand{\argmin}{\operatornamewithlimits{argmin}}

     \newcommand{\Rmnum}[1]{\expandafter\@slowromancap\romannumeral #1@}

\title{Uncertainty Quantification for Data-Driven Change-Point Learning via Cross-Validation}
\author{
 Hui Chen\textsuperscript{\rm 1}\footnote{These authors contributed equally, and are listed in alphabetical order.}
 Yinxu Jia\textsuperscript{\rm 2}\footnotemark[1], 
 Guanghui Wang\textsuperscript{\rm 3}\footnotemark[1]\thanks{Guanghui Wang is the corresponding author.}, 
 Changliang Zou\textsuperscript{\rm 2}\footnotemark[1]
}
\affiliations{
    \textsuperscript{\rm 1}School of Mathematics and Statistics, Jiangsu Normal University\\
    \textsuperscript{\rm 2}School of Statistics and Data Science, LPMC, KLMDASR, and LEBPS, Nankai University\\
    \textsuperscript{\rm 3}School of Statistics, Academy of Statistics and Interdisciplinary Sciences, and KLATASDS-MOE, East China Normal University \\
    chenhuiah@163.com,  yxjia@mail.nankai.edu.cn,  ghwang.nk@gmail.com,  zoucl@nankai.edu.cn
}

\begin{document}

\maketitle

\begin{abstract}
Accurately detecting multiple change-points is critical for various applications, but determining the optimal number of change-points remains a challenge. Existing approaches based on information criteria attempt to balance goodness-of-fit and model complexity, but their performance varies depending on the model. Recently, data-driven selection criteria based on cross-validation has been proposed, but these methods can be prone to slight overfitting in finite samples. In this paper, we introduce a method that controls the probability of overestimation and provides uncertainty quantification for learning multiple change-points via cross-validation. We frame this problem as a sequence of model comparison problems and leverage high-dimensional inferential procedures. We demonstrate the effectiveness of our approach through experiments on finite-sample data, showing superior uncertainty quantification for overestimation compared to existing methods. Our approach has broad applicability and can be used in diverse change-point models.
\end{abstract}

\section{Introduction}
Change-point detection plays a crucial role in diverse domains, including machine learning and statistics \cite{Aue:2013hg, Niu:2016et, aminikhanghahi2017survey, fearnhead2020relating, truong2020selective, cho2021data}. Consider a scenario where we have collected a sequence of independent data observations from a parametric multiple change-point model. In this model, the data is divided into \(K_{n}+1\) pieces, each adhering to a parametric model. Importantly, the parameters differ between successive pieces, representing variations such as the mean or variance of the data, or regression coefficients illustrating the relationship between a response and covariates.

 Among the primary challenges in the multiple change-point model is the estimation of the number of change-points. Existing approaches involve selecting an optimal threshold or penalty, where the ideal value may depend on the specific model at hand. For instance, in binary segmentation and its variants \citep{venkatraman1992consistency, Fryzlewicz:2014dh,harchaoui2008kernel, li2015m}, each step of the partitioning algorithm requires the determination of a threshold to decide when to stop segment partitioning. Nonetheless, identifying the appropriate threshold poses a significant challenge. Similarly, in the case of penalized minimization algorithms \citep{Yao:1988jk, Bai:1998hp, Braun:2000de, Hannart:2012iw, killick2012optimal, Zou:2014bn, haynes2017computationally, haynes2017, Wang:2017vm, cho2022two, baranowski2019narrowest, fearnhead2019detecting}, it becomes crucial to carefully select the penalized parameters.

In recent developments, \citet{zou2020consistent} introduced a data-driven selection criterion utilizing cross-validation (CV) to address the issue of threshold or penalty selection. This criterion is versatile, as it can be applied alongside various change-point detection algorithms and is suitable for different parametric change-point models. Building upon this work, \citet{pein2021cross} further refined the cross-validation criterion with a specific focus on detecting changes that involve large magnitudes.

Cross-validation has been widely used as a technique to estimate the prediction error of a model \citep{allen1974relationship, stone1974cross, geisser1975predictive}. The fundamental concept behind cross-validation is to fit and evaluate candidate models on separate datasets to obtain an unbiased performance evaluation. However, traditional cross-validation methods, such as leave-one-out and V-fold variants, have limitations when it comes to model selection. These methods can suffer from overfitting, which leads to inaccurate model selection. Theoretical studies by \citet{shao1993linear, zhang1993model, yang2007consistency} have shown that cross-validation may not consistently select the correct model in low-dimensional linear models unless the training-testing split ratio tends to zero. Unfortunately, achieving such ideal split ratios is not useful in real-world applications. Recent advances in this field by \citet{austern2020asymptotics, bayle2020cross} have focused on studying the asymptotic distribution of the cross-validated risk under certain stability conditions.

The cross-validation criterion in the context of change-point detection problems introduced by \citet{zou2020consistent} demonstrated that the criterion can lead to consistent selection of the number of change-points. However, achieving consistency between the estimated number of change-points, denoted as $\widehat{K}_{\rm CV}$, and the true $K_n$ may require stringent conditions for the change-point detection algorithms. In practical applications, this criterion can be susceptible to slight overfitting. To gain further insights into this phenomenon, Table \ref{CVOE} provides a measure of $\widehat{K}_{\rm CV}-K_n$ for two commonly used change-point detection algorithms: wild binary segmentation (WBS) \citep{Fryzlewicz:2014dh} and the pruned exact linear time algorithm (PELT) \citep{killick2012optimal}. The results indicate a slight overestimation, which can be beneficial as it helps capture important change signals, preventing them from being missed.

The main objective of this paper is to assess the level of uncertainty associated with cross-validation in change-point detection. We are fortunate that the slight overestimation of the cross-validation criterion can be measured and managed in real-world scenarios, with theoretical assurance under less stringent conditions compared to achieving the consistency of $\widehat{K}_{\rm CV}$. This research accomplishes this by showcasing that the likelihood of overestimation while utilizing the cross-validation criterion can be adequately controlled. Through these findings, we contribute to a better understanding of the cross-validation criterion in change-point detection.

In recent years, there has been a growing interest in quantifying the uncertainty associated with the number of change-points. \citet{frick2014multiscale} proposed a methodology based on multi-scale statistics to detect change-points while controlling the probability of overestimating at least one change-point. Expanding on this approach, \citet{lim2016} refined this methodology by incorporating false discovery rate control. More recently, \citet{chen2021data} proposed a data-driven approach that leverages globally symmetric statistics to achieve false discovery rate control. These approaches quantify uncertainty for the number of change-points via error rate control in multiple testing framework.
On the other hand, the method in \citet{frick2014multiscale} is distribution-dependent and cannot keep the probability of overestimation in the case of more complex data.

Our work is inspired by the ``cross-validation with confidence" method recently introduced by \citet{lei2020cross}. Their method incorporates cross-validation to establish a framework for quantifying the uncertainty related to selected models or tuning parameters. In a similar vein, our work aims to provide a measure of uncertainty for cross-validation estimates in the context of change-point detection.

We focus on effectively quantifying the difference $\widehat{K}_{\rm CV}-K_n$ with a certain confidence level. The key idea is to find a lower bound, $\widehat{K}_{\rm min}$, for $K_n$ that provides confidence. To control the overestimation of $\widehat{K}_{\rm CV}$, we can utilize the difference, $\widehat{K}_{\rm CV}-\widehat{K}_{\rm min}$. In order to find $\widehat{K}_{\rm min}$, we conduct a sequence of hypothesis tests for each candidate number of change-points. The null hypothesis is that a specified model estimated from a candidate $r$ has the smallest prediction error among the candidates larger than $r$. This test is conducted sequentially and individually for each candidate until we accept the null hypothesis for the first time. To obtain a valid critical value, we implement a high-dimensional mean testing procedure that tests for prediction error discrepancy for a new sample.

Table \ref{CVOE} presents the average values of $\widehat{K}_{\rm CV}-K_n$ and $U=\widehat{K}_{\rm CV}-\widehat{K}_{\rm min}$, and the empirical overestimation probability $P_+:=\mathbb{P}(\widehat{K}_{\rm CV}-K_n>U)$ based on $500$ replications. The results are obtained for two change-point detection algorithms WBS and PELT, across different sample sizes $n$. It is observed that the data-driven cross-validation estimate $\widehat{K}_{\rm CV}$ indicates slight overestimation and we can control the overestimation.

\begin{table}[t]      \tabcolsep 2pt
\centering
% \linespread{1.3}
{\small \centering\begin{tabular}{ccccccccccccccccccccccccccc}
\toprule
& &&$n$\\
Algorithm   &Measure                                            & $300$&$400$&$500$&$600$& $700$&$800$\\
\hline	
 WBS  	     &	$\widehat{K}_{\rm CV}-K_n$	                     &0.7  &0.8   &0.5&0.5 &0.4&0.4   \\
   &	$U$	 &0.9  &0.9   &0.5&0.5 &0.4&0.4 \\
         &  $P_+$ 	                                             &2.0  &1.0   &1.5&1.5 &1.5&1.0\\
\\
PELT    	 &	$\widehat{K}_{\rm CV}-K_n$	                     &0.4  &0.3   &0.3&0.3 &0.3&0.3 \\
&$U$	 &0.5  &0.4   &0.4&0.3 &0.3&0.3\\
         &  $P_+$ 	                                             &1.0  &1.0   &0.0&0.0 &0.0&0.0\\
\bottomrule

\end{tabular}}
\caption{{{
Average of $\widehat{K}_{\rm CV}-K_n$ and $U=\widehat{K}_{\rm CV}-\widehat{K}_{\rm min}>0$, and empirical overestimation probability $P_+:= \mathbb{P}(\widehat{K}_{\rm CV}-K_n>U)$ (in \%), for two change-point detection algorithms WBS and PELT, across different sample sizes $n$.
The nominal level is $\alpha=10\%$. The data is generated from Gaussian distribution with a signal-to-noise ratio $1$ and $K_n=5$; see Section \ref{simulation}.}}} \label{CVOE}
\end{table}

The main contributions and advantages of our proposal are as follows:
\begin{itemize}
\item Uncertainty quantification for cross-validation criterion: Our proposal quantifies the uncertainty of cross-validation in change-point detection \citep{zou2020consistent}. By utilizing hypothesis testing and model comparison techniques, we can control the probability of overestimation within a specified level.
     %This allows researchers and practitioners to have a better understanding of the reliability of cross-validation in change-point learning.

\item Weaker requirements for control: Our research demonstrates that achieving control over the probability of overestimation can be achieved with weaker requirements on the change-point detection algorithms compared to those needed for consistency, as evidenced by the work of \citet{zou2020consistent}. This finding highlights the robustness and reliability of the cross-validation criterion in change-point detection.

\item Versatility: Our method is versatile and can be used in conjunction with various change-point detection algorithms. It is also suitable for different change-point models, making it adaptable to a wide range of applications and scenarios.
\end{itemize}

Notation. Denote ${\bm 1}(\cdot)$ as the indicator function. 
Let $\{{\bf x}_1,\ldots,{\bf x}_n\}$ and $\{{\bf y}_1,\ldots,{\bf y}_n\}$ be two sets of $d$-variate vectors. Define the norm $\|{\bf x}\|=\sqrt{{\bf x}^\top {\bf x}}$. For any interval $(a,b]$ with $0\leq a < b \leq n$, denote $\bar{\bf x}_{a,b}=(b-a)^{-1}\sum_{i=a+1}^b{\bf x}_i$. Let $\mathcal{T}_r=(\tau_1,\ldots,\tau_r)$ be a set of $r$ points such that $0<\tau_1<\cdots<\tau_r<n$.
We denote
\begin{align*}
    \mathcal{C}_{{\bf x}{\bf y}}(\mathcal{T}_r)=&\sum_{j=0}^r(\tau_{j+1}-\tau_j)\bar{\bf x}^{\top}_{\tau_j, \tau_{j+1}}\bar{\bf y}_{\tau_j, \tau_{j+1}}
\end{align*}
where $\tau_0=0$ and $\tau_{r+1}=n$. Define  $\mathcal{C}_{\bf x}^2=\mathcal{C}_{{\bf x}{\bf x}}$.
\section{Methodology}\label{method}
\subsection{The problem}
Suppose we have a sequence of independent data observations $\bm\xi=\{\bm\xi_1,\ldots,\bm\xi_n\}$ from a parametric
multiple change-point model:
\begin{align}\label{model:general}
\bm\xi_{i}\sim \mathcal{G}(\cdot\mid\bm\beta^*_k), ~ \tau_{k}^* <i \leq \tau_{k+1}^*,~k=0,\ldots,K_n; \ i=1,\ldots,n,
\end{align}
where $K_n$ is the true number of change-points, $\tau_k^*$'s are the locations of change-points with the convention of $\tau_0^*=0$ and $\tau_{K_n+1}^*=n$, $\bm\beta^*_k$ is a $d$-dimensional parameter vector of interest and $\mathcal{G}(\cdot\mid\bm\beta^*_k)$ represents the model structure of the segment $k$ satisfying $\bm\beta^*_k\neq\bm\beta^*_{k+1}$.
In this model, we use the notation $\bm\xi_i$ which could either be a $d$-variate random vector or $(Y_i,\X_i)$ with $Y_{i}$ and $\X_{i}$ being respectively the response variable and $d$-variate explanatory variable.

{\bf Problem} (Uncertainty quantification for
change-point detection via cross-validation)
\begin{enumerate}
\item[] Learn the change-points specific to the task at hand by employing the cross-validation criterion proposed by \citet{zou2020consistent}, and obtain an estimated number of change-points $\widehat{K}_{\rm CV}$;
\item[] Aim to demonstrate that the probability of overestimation, defined as $\mathbb{P}(\widehat{K}_{\rm CV}-K_n>U)$, can be effectively controlled under a prescribed level $0<\alpha<1$. Here $U$ is a quantity to be determined.
\end{enumerate}

Selecting an appropriate value for $U$ is crucial to effectively control the level of overestimation. We approach this problem by formulating a series of hypothesis testing problems as follows:
\begin{align}\label{K}
H_{0,r}:K_n= r,~~H_{1,r}:K_n> r,\ 0\leq r\leq p_n,
\end{align}
where $p_n$ represents an upper bound on the number of change-points.
We initiate the process with $r=0$ and test the hypothesis (\ref{K}) using the desired level $\alpha$. If we reject $H_{0,r}$, we increment $r$ by 1 and proceed to the next hypothesis. This iterative process continues until we accept $H_{0,r}$ for the first time. The resulting value of $r$ is denoted as $\widehat{K}_{\rm min}$. We will show that, by choosing $U=\widehat{K}_{\rm CV}-\widehat{K}_{\rm min}\ge 0$,
\begin{align*}
\mathbb{P}(\widehat{K}_{\rm CV}-K_n>U)\leq\alpha.
\end{align*}

The null hypothesis being tested is that the true number of change-points in the model is equal to $r$. To test this hypothesis, we compare the prediction error discrepancy between a model with $r$ change-points and other models with $s$ change-points with $s>r$. The prediction error discrepancy is denoted as $\Delta_{r,s}$.

If $r$ represents the optimal number of change-points, we expect the model with $r$ change-points to have a lower prediction error compared to models with a larger number of change-points. In other words, we anticipate that the maximum value of $\Delta_{r,s}$ when $s > r$ will be less than or equal to 0.

 Formally, we consider the following hypothesis:
\begin{align}\label{test}
H_{0,r}:\max_{s> r}\Delta_{r,s}\leq 0,~~H_{1,r}:\max_{s> r}\Delta_{r,s}> 0.
\end{align}
This testing problem in fact involves a sequence of $(p_n-r-1)$-dimensional vectors $(\Delta_{r,s}: s>r)$, for $r=0,1,\ldots$, where $p_n$ can diverge to infinity as the sample size $n\to\infty$. Hence some high-dimensional inferential tools may be leveraged for designing a valid testing procedure.

If the null hypothesis $H_{0,r}$ is rejected, it suggests that adding more change-points to the model could potentially improve prediction performance. In such a case, it becomes necessary to consider models with a greater number of change-points for a more accurate prediction.

\subsection{Testing procedure}\label{subsec:ss-procedure}
Consider the general parametric model (\ref{model:general}). To estimate the underlying model and evaluate the prediction error based on different data,
we employ a sample-splitting strategy. That is, we use one part of the data to train the model and another part to compute the prediction error.
Inspired by the order-preserved splitting strategy proposed by \citet{zou2020consistent}, we partition the data into training set
$\bm\xi^{\rm tr}$ and validation set $\bm\xi^{\rm te}$ with $n_{\rm tr}$ and $n_{\rm te}$ observations respectively, according to whether the observed index being odd or even, such that the two data sets share a similar change-point pattern as much as possible.

Next, we can apply some commonly used change-point detection algorithm
${\mathcal{A}(\cdot)}$ to learn $r$ change-points based on the data set $\bm\xi^{\rm tr}$, and obtain the set of change-points $\Th_{r}=(\th_{r,1},\ldots,\th_{r,r})$. In the meantime, we obtain the estimated parameters, i.e., $\widehat{\mathcal{B}}=\{\widehat{\bm\beta}_0,\ldots,\widehat{\bm\beta}_{r}\}$. The resulting model
is denoted as $\widehat{\mathcal{U}}^{\rm tr}_r=\{\Th_{r},\widehat{\mathcal{B}}\}$. Correspondingly, for $\bm\xi_i\in\bm\xi^{\rm te}$, we introduce the prediction error
$\mathbb{E}\{\mathcal{O}(\bm\xi_{i};\widehat{\mathcal{U}}^{\rm tr}_r)\mid{\bm\xi}^{\rm tr}\}$, where
\begin{align*}
\mathcal{O}(\bm\xi_{i};\widehat{\mathcal{U}}^{\rm tr}_r)=\sum_{j=0}^r\ell(\bm\xi_i,\widehat{\bm\beta}_j){\bf 1}(\th_{r,j}<i\leq \th_{r,j+1}),
\end{align*}
here $\ell(\cdot,\cdot)$ is a loss function. Take the classic quadratic loss as an example. Under the mean shift model or the linear regression model with structural breaks, we have $\ell(\bm\xi_i,\widehat{\bm\beta}_j)=\|\bm\xi_i-\widehat{\bm \beta}_j\|^2$
or $\ell(\bm\xi_i,\widehat{\bm\beta}_j)=(Y_i-\X_i^{\top}\widehat{\bm\beta}_j)^2$, respectively. Our method can be used beyond the quadratic loss, such as the quantile loss or other robust loss functions.

For any $s> r$, define the discrepancy of the loss
for $\bm\xi_i$ between Model $r$ and $s$ by
\begin{align*}
\delta^{(i)}_{r,s}=\mathcal{O}(\bm\xi_{i};\widehat{\mathcal{U}}^{\rm tr}_r)-\mathcal{O}(\bm\xi_{i};\widehat{\mathcal{U}}^{\rm tr}_s),~ i\in\mathcal{I}_{\rm te},
\end{align*}
 where $\mathcal{I}_{\rm te}$ is the index set of $\bm\xi^{\rm te}$. Correspondingly, we define
the discrepancy of the prediction error between Model $r$ and $s$, $\Delta_{r,s}=\sum_{i\in\mathcal{I}_{\rm te}}\mathbb{E}(\delta^{(i)}_{r,s}\mid\bm\xi^{\rm tr})/n_{\rm te}$.

We adopt the test statistic
\begin{align*}
\mathcal{D}=\max_{s> r}\sqrt{n_{\rm te}}{\widehat{\Delta}_{r,s}}/{\widehat{\sigma}_{r,s}},
\end{align*}
where $\widehat{\Delta}_{r,s}$ and  $\widehat{\sigma}^2_{r,s}$ are the sample mean and variance of the sequence $\{\delta^{(i)}_{r,s}:i\in\mathcal{I}_{\rm te}\}$, respectively.

To approximate the null distribution of $\mathcal{D}$, we use a multiplier bootstrap approach. For $b=1,\ldots,B$, we generate independent and identically distributed (iid) random variables $e_i\in N(0,1)$, for $i\in \mathcal{I}_{\rm te}$. Let
\begin{align*}
\mathcal{D}_b=\max_{s> r}n_{\rm te}^{-1/2}\sum_{i\in\mathcal{I}_{\rm te}}(\delta^{(i)}_{r,s}-\widehat{\Delta}_{r,s})e_i/{\widehat{\sigma}_{r,s}}.
\end{align*}
We obtain the critical value
\begin{align}\label{th}
c_{r,\alpha}=\inf\left\{t\in \mathbb{R}:B^{-1}\sum_{b=1}^B{\bf 1}(\mathcal{D}_b> t)\leq \alpha\right\},
\end{align}
and then reject $H_{0,r}$ if $\mathcal{D}>c_{r,\alpha}$. If we reject $H_{0,r}$, we increment $r$ by 1 and proceed to the next hypothesis. This iterative process continues until we accept $H_{0,r}$ for the first time. We denote   $\widehat{K}_{\rm min} = \min\{r: H_{0, r} \text{ is accepted}\}$.

\subsection{Refinement with cross-validation}\label{sectionCV}

To improve the stability and mitigate the power loss due to sample-splitting, we introduce the V-fold cross-validation. Define the discrepancy of the loss
for $\bm\xi_i$ between Model $r$ and $s$ by
\begin{align*}
\delta^{(i)}_{r,s}=\mathcal{O}(\bm\xi_{i};\widehat{\mathcal{U}}^{(-v)}_r)-\mathcal{O}(\bm\xi_{i};\widehat{\mathcal{U}}^{(-v)}_s),~ i\in\mathcal{I}_{v},~v=1,\ldots,V,
\end{align*}
where $\mathcal{I}_{v}$ is the index set of the $v$-th fold, and $\widehat{\mathcal{U}}^{(-v)}_r$ denotes the model fitted by the data $\bm\xi^{(-v)}$ with
the $v$-th fold removing from the full data.
Accordingly, the discrepancy of the prediction error between Model $r$ and $s$ for the $v$-th fold is
\begin{align*}
\Delta^{(v)}_{r,s}=n^{-1}V\sum_{i\in\mathcal{I}_{v}}\mathbb{E}(\delta^{(i)}_{r,s}\mid\bm\xi^{(-v)}).
\end{align*}
Denote $\Delta_{r,s}=V^{-1}\sum_{v=1}^V\Delta^{(v)}_{r,s}$. We propose the following steps to implement the hypothesis testing for (\ref{test}).
To construct the test statistic, we introduce the estimate of $\Delta_{r,s}$,
$\widehat{\Delta}_{r,s}=V^{-1}\sum_{v=1}^V\widehat{\Delta}^{(v)}_{r,s}$,
 here
$\widehat{\Delta}^{(v)}_{r,s}=n^{-1}V\sum_{i\in\mathcal{I}_{v}}\delta^{(i)}_{r,s}$.

We denote the group-wise centered discrepancy of predictive loss between Model $r$ and $s$ by $\tilde{\delta}^{(i)}_{r,s}=\delta^{(i)}_{r,s}-\widehat{\Delta}^{(v)}_{r,s},~i\in \mathcal{I}_v$.
Let $\widehat{\sigma}_{r,s}$ be the sample standard deviation of $\{\tilde{\delta}_{r,s}^{(i)}: 1\leq i\leq n\}$.
 We propose the statistic
\begin{align*}
\tilde{\mathcal{D}}=\max_{s> r}\sqrt{n}{\widehat{\Delta}_{r,s}}/{\widehat{\sigma}_{r,s}}.
\end{align*}
Then we reject $H_{0,r}$ if $\tilde{\mathcal{D}}>c_{r,\alpha}$, where $c_{r,\alpha}$ is the critical value of the test.

To derive $c_{r,\alpha}$, we also generate iid standard Gaussian random variables $e_i$, $1\leq i \leq n$, accordingly
\begin{align*}
\tilde{\mathcal{D}}_b=\max_{s> r}n^{-1/2}\sum_{v=1}^V\sum_{i\in\mathcal{I}_{v}}({\delta^{(i)}_{r,s}-\widehat{\Delta}^{(v)}_{r,s}})e_i/{\widehat{\sigma}_{r,s}}
\end{align*}
for $b=1,\ldots,B$. We obtain the critical value $c_{r,\alpha}$ according to (\ref{th}) with $\mathcal{D}_b$ replaced by $\tilde{\mathcal{D}}_b$.
\section{Theoretical analysis}

In this section, we present theoretical results of our method under the  mean  shift model. That is, $d$-variate observations $\bm\xi_i$'s follow
\begin{align*}
\bm \xi_i={\bm \mu}_i+\bm\varepsilon_i,~~i=1,\ldots,n,
\end{align*}
where ${\bm\varepsilon}_i$ is a $d$-dimensional random vector with mean zero, and $\bm\mu$ is piecewise constant with change-points $\tau_1^*<\dots<\tau_{K_n}^*$; that is $\bm\mu_i\neq \bm\mu_{i+1}$ if and only if $i=\tau_k^*$ for some $k$ and $i=1,\ldots, n-1$. For $k=0,\ldots,K_n$, ${\bm\beta}^*_k=\mathbb{E}({\bm\xi}_i)$, $\bm\Sigma_{k}={\rm cov}({\bm\varepsilon}_i)$ for $\tau_k^*<i\leq \tau_{k+1}^*$.  Denote the maximum and minimum eigenvalues of $\bm\Sigma_k$ by $\overline{\sigma}(\bm\Sigma_k)$ and $\underline{\sigma}(\bm\Sigma_k)$ for $k=0,\ldots,K_n$. Let
$\overline{\sigma}=\max\{\overline{\sigma}(\bm\Sigma_0),\ldots,\overline{\sigma}(\bm\Sigma_{K_n})\}$ and $\underline{\sigma}={\rm min}\{\underline{\sigma}(\bm\Sigma_0),\ldots,\underline{\sigma}(\bm\Sigma_{K_n})\}$. The maximal and minimal distance between change points are denoted as $\overline{\lambda}_n=\max_{0\leq k\leq {K_n}}(\tau_{k+1}^*-\tau_{k}^*)$ and $\underline{\lambda}_n=\min_{0\leq k\leq {K_n}}(\tau_{k+1}^*-\tau_{k}^*)$. Let
 $\gamma_n^{(k)}=\|\bm\beta^*_{k-1}-\bm\beta^*_{k}\|^2$ for $1\leq k\leq {K_n}$ and
the minimal signal strength be $\underline{\gamma}_n=\min_{1\leq k\leq {K_n}}\|\bm\beta^*_{k-1}-\bm\beta^*_{k}\|^2$.
Without loss of generality, we use the order-preserved splitting strategy to split the data based on odd and even indices. For convenience, assume that the sample size $n$ is an even number.  We denote
\begin{align*}
&\bm\xi_i^{\rm tr}=\bm\mu_i^{\rm  tr}+\bm\varepsilon^{\rm tr}_i,~i\in\mathcal{I}_{\rm tr},~\bm\xi_i^{\rm te}=\bm\mu_i^{\rm te}+\bm\varepsilon^{\rm te}_i,~i\in\mathcal{I}_{\rm te}.
\end{align*}
 The quadratic loss function for the model $\widehat{\mathcal{U}}_r^{\rm tr}$ for $j=0,\ldots, r$ and $\widehat{\tau}_{r, j}<i\leq \widehat{\tau}_{r, j+1}$ is of the form
\begin{align*}
\ell(\bm\xi_{i}^{\rm te},\widehat{\bm\beta}_j)=\|\bm\xi_{i}^{\rm te}-\bar{\bm\xi}_{\widehat{\tau}_{r, j}, \widehat{\tau}_{r, j+1}}^{\rm tr}\|^2.
\end{align*}
For other change-points models, we can use scores and transform the problem to the mean domain \citep{zou2020consistent}.

Let $\Delta_{r, s}=\mathbb{E}(\widehat{\Delta}_{r, s}\mid{\bm\xi}^{\rm tr})$ and  $\sigma_{r, s}^2=\mathbb{E}(\widehat{\sigma}_{r, s}^2\mid {{\bm\xi}}^{\mathrm{tr}})$, where $\widehat{\Delta}_{r, s}$ and $\widehat{\sigma}_{r, s}^2$ have been defined in Section~\ref{subsec:ss-procedure}.   Let $\eta_{r, s}^{(i)}=\sigma_{r, s}^{-1}(\delta_{r, s}^{(i)}-\Delta_{r, s})$ for $i\in\mathcal{I}_{\rm te}$ and $0\leq r<s\leq p_n$.

We need the following assumption to facilitate our theoretical demonstration.
\begin{assu}[{\it Moments}]\label{assu:seup}
(i) There exists some $\vartheta>2$ such that $\mathbb{E}(|\eta_{r, s}^{(i)}|^\vartheta)<\infty$ for every $i\in\mathcal{I}_{\mathrm{te}}$;

(ii) There exists some some constants $\nu\in(0,1/2$) and $C>0$ such that
\begin{align}\label{assu1}
(M_{n,3}^3\vee M_{n,4}^2\vee B_{n})^2\log^{7/2} p_n\leq Cn^{1/2-\nu},
\end{align}
where $M_{n, k}=\max_{i,  r}\max_{s>r}\mathbb{E}^{1/k}(|\eta_{r, s}^{(i)}|^k)$ for $k=3, 4$ and $B_n=\max_{i, r}\mathbb{E}^{1/4}\{\max_{s>r}(\eta_{r, s}^{(i)})^4\}$.
\end{assu}
Assumption~\ref{assu:seup} is motivated by \citet{lei2020cross} and puts restrictions on the moments of $\delta_{r, s}^{(i)}$. \citet{lei2020cross} in fact assumed sub-exponential tail conditions. Both kinds of conditions  are common in developing high-dimensional central limit theorems and justifying the validity of the procedures; see \citet{chernozhukov2017central,chernozhukov2019inference} for details.
\begin{thm}\label{thmconsis} Suppose that Assumptions \ref{assu:seup} holds and for $s>K_n$,
\begin{align}\label{equ:variation}
         \{\mathcal{C}_{\bm\xi^{\rm tr}}^2(\widehat{\mathcal{T}}_{K_n})-\mathcal{C}_{\bm\xi^{\rm tr}}^2(\widehat{\mathcal{T}}_s)\} - 2\{\mathcal{C}_{\bm\mu^{\rm tr}\bm\xi^{\rm tr}}(\widehat{\mathcal{T}}_{K_n})-\mathcal{C}_{\bm\mu^{\rm tr}\bm\xi^{\rm tr}}(\widehat{\mathcal{T}}_{s})\}\leq0
\end{align}
holds with probability one.
 Then $\mathbb{P}(  \widehat{K}_{\rm CV}-K_n>  U)\leq \alpha+o(1)$.
\end{thm}
\remark 1
    Theorem~\ref{thmconsis} shows that the probability of overestimation $\mathbb{P}(\widehat{K}_{\rm CV}-K_n>U)$ can be effectively controlled for the sample-splitting procedure. The main step of the proof is to show that the hypothesis $H_{0, r}$ with the true value of $r=K_n$ will not be rejected. Condition (\ref{equ:variation}) puts some requirements on the change-point detection algorithm when the model is overfitting. It is weaker than Condition (10) imposed in \citet{zou2020consistent} to achieve consistent selection of the number of change-points.
    Since we are testing the discrepancy of two prediction errors conditional on the training set by (\ref{test}), our method eliminates the randomness from the testing set. On the contrary, the cross-validation procedure to render consistency requires that the randomness of the training set overrides the randomness from the testing set \cite{zou2020consistent}. To see this, consider  the univariate sequence with $K_n=0$ (no change-point). Suppose that the number of change-points is overestimated and $s=1$. In this case, it can be verified that the second term of (\ref{equ:variation}) is $0$. Some simple algebra shows that the first term  of (\ref{equ:variation}) is of the form
\begin{align*}
    \mathcal{C}_{\bm\xi^{\rm tr}}^2(\widehat{\mathcal{T}}_{0})-\mathcal{C}_{\bm\xi^{\rm tr}}^2(\widehat{\mathcal{T}}_1)=-\frac{\widehat{\tau}(n_{\rm tr}-\widehat{\tau})}{n_{\rm tr}}(\bar{\bm\xi}_{0, \widehat{\tau}}^{\rm tr}-\bar{\bm\xi}_{\widehat{\tau}, n_{\rm tr}}^{\rm tr})^2\leq 0,
\end{align*}
where $\widehat{\tau}$ is the estimated location of the change-point under $s=1$. Hence, Condition (\ref{equ:variation}) holds immediately.
Furthermore, assume there exists one true change-point $\tau^*$. Suppose we identify a change-point at $\widehat{\tau}$ for $s=1$, and find two change-points $\widehat{\tau}_1$ and $\widehat{\tau}_2$ with $\widehat{\tau}_1<\widehat{\tau}_2$ for $s=2$.
Consider $\tau^*\leq \widehat{\tau}\leq \widehat{\tau}_1$. Condition (\ref{equ:variation}) holds by observing that
\begin{align*}
(\widehat{\tau}_1^{-1}-\widehat{\tau}^{-1})\{\gamma_n^{(1)}\}^2+
    \mathcal{C}_{\bm\varepsilon^{\rm tr}}^2(\widehat{\mathcal{T}}_1)-\mathcal{C}_{\bm\varepsilon^{\rm tr}}^2(\widehat{\mathcal{T}}_2)\leq 0.
\end{align*}
In contrast, to achieve consistency, Condition (10) in \citet{zou2020consistent} generally requires that $-\{\mathcal{C}_{\bm\xi^{\rm tr}}^2(\widehat{\mathcal{T}}_{K_n})-\mathcal{C}_{\bm\xi^{\rm tr}}^2(\widehat{\mathcal{T}}_{s})\}$ is of larger order than $\log\log n$.

\begin{pro}\label{prop:underestimation}
    (i) If $\alpha<0.5$, $ \lim_{B\to\infty}\mathbb{P}(\widehat{K}_{\rm CV}\geq \widehat{K}_{\rm min})=1$.

    (ii) Suppose that the $\bm\Sigma_j$'s are positive-definite matrices and there exists a positive integer $m\geq 2$ such that $\mathbb{E}(\|\bm\Sigma_j^{-1/2}\bm\varepsilon_i\|^{2m})<\infty$ for $i\in(\tau^*_j, \tau_{j+1}^*]$, $j=0, \ldots, K_n$.  If the jump sizes $\|\bm\beta^*_{k-1}-\bm\beta^*_{k}\|$'s satisfy
\begin{align}\label{equ:underfit}
    \frac{\underline{\lambda}_n\underline{\gamma}_n}{{K_n}\overline{\sigma}\overline{\lambda}_n^{2/m}}&\rightarrow\infty,
\end{align}
then we have $\lim_{n\to\infty}\mathbb{P}(\widehat{K}_{\rm CV}\geq K_n)=1$.
\end{pro}

Proposition~\ref{prop:underestimation} ensures that $U\geq 0$ and $\widehat{K}_{\rm CV}-K_n\geq 0$, which makes our control of overestimation probability more useful. Assumptions in Proposition~\ref{prop:underestimation}--(ii) are from \citet{zou2020consistent}, which put some requirements on the minimal signal strength and distance between two change-points to avoid underestimation.
\section{Experiments}
In this section, we demonstrate the performance of the proposed methods through numerical studies. 
The experiments are run on a personal computer with an Intel Core i7-10700 CPU, 8GB of memory, a 64-bit operating system and R software version 4.2.1.
\subsection{Synthetic data}\label{simulation}
{\bf Data}. Take the model $\xi_i=\mu_i+\sigma\varepsilon_i$ as an example. We set the signal vector $\bm\theta=(-1,1)$,
and let $\theta_{k-1}$ denote the $k$-th element of $\bm\theta$. Take $\mu_i=\theta_{{\rm mod}(\mathcal{J}_0+j,2)}$ for $\tau_j^*<i\leq\tau_{j+1}^*,j=0,\ldots,K_n$, where $\mathcal{J}_0$ is an integer randomly sampling from $\{1,2\}$ and ${\rm mod}(~,~)$ is the modulo operator. The signal-to-noise (SNR) is defined as $\text{sd}(\mu_i\text{'s})/\sigma$, where $\text{sd}(\mu_i\text{'s})$ is the standard deviation of the signals $\{\mu_1,\ldots,\mu_n\}$.

The error terms (i) $\varepsilon_i\overset{\rm iid}{\sim}N(0,1)$; (ii) $\varepsilon_i\overset{\rm iid}{\sim}\sqrt{0.6}t(5)$, $i=1,\ldots,n$ are considered, where $t(5)$ is $t$ distribution
with $5$ degrees of freedom.
The change-points $\tau_j^*=j\lfloor n/(K_n+1)\rfloor+{\rm Uniform}(-a,a)$ with $a=\lfloor n^{1/4}\rfloor$ for $j=1,\ldots,K_n$, where ${\rm Uniform}(a,b)$ is the continuous uniform distribution with support $[a,b]$. All the simulation results are based on $500$ replications and the bootstrap sample size is $B=500$.

{\bf Methods}. In the experiments, we evaluate the performance of our sample-splitting-based method (SS) for quantifying the uncertainty of overestimation of cross-validation criterion. We also investigate the refinement of our method using multiple folds cross-validation (RC) as discussed in Section \ref{sectionCV}.
For illustration purposes, we conduct the experiments using a three-fold cross-validation approach.

In this study, we examine the cross-validation criterion in \citet{zou2020consistent} in combination with the WBS and PELT algorithms. We use the implementations of these algorithms provided in the R packages "wbs" \citep{wbs} and "changepoint" \citep{killick2016changepoint,killick2014changepoint}.
The method we propose, which utilizes sample-splitting along with the PELT algorithm, is referred to as SS-PELT. The other methods are named following a similar convention.
The approach introduced by \citet{frick2014multiscale} (abbreviated as SMUCE) is taken into account as a benchmark method. SMUCE is implemented in the R package \texttt{stepR} \citep{SMUCE}.

{\bf Performance measures}.
 A key metric is the empirical overestimation probability \(P_+\). It is essential for \(P_+\) to be lower than the specified nominal level \(\alpha\), demonstrating controlled overestimation. Additionally, we will emphasize the average values of \(\widehat{K}-K_n\), which should be slightly larger than 0 (with smaller values being more desirable), where $\widehat{K}$ denotes the estimates $\widehat{K}_{\rm SMUCE}$ and $\widehat{K}_{\rm CV}$
for SMUCE and CV. This emphasizes that \(\widehat{K}\) is a reliable estimator of the true number of change-points \(K_n\) and prevents underestimation.

{\bf Results}.
\begin{table}[ht]      \tabcolsep 1.5pt
\centering
\linespread{1.3}
{\small \centering\begin{tabular}{clllccccccccccccccccccccccc}
\toprule
&&& \multicolumn{2}{c}{$\alpha=5\%$} && \multicolumn{2}{c}{$\alpha=10\%$}&& \multicolumn{2}{c}{$\alpha=20\%$}\\
\cline{4-5}\cline{7-8} \cline{10-11}
Error&$K_n$&Method & $P_+$&$\widehat{K}-K_n$&&$P_+$&$\widehat{K}-K_n$&& $P_+$&$\widehat{K}-K_n$\\
\hline	
$N(0,1)$ &25	     &	SMUCE	 &0.0  &0.0          &&0.0&0.0          &&0.0&0.0   \\
         &           &	SS-PELT	 &0.0  &0.6    &&1.0&0.6    &&3.0&0.6 \\  	
         &35         &	SMUCE	 &0.0  &-0.2         &&0.0&-0.1          &&1.0&-0.1   \\
         &      	 &	SS-PELT	 &1.0  &0.5    &&2.0&0.5      &&3.0&0.5   \\
\\
$t(5)$   &25	     &	SMUCE	 &63.5  &1.8        &&77.5&2.4          &&83.0&3.5   \\
         & 	         &	SS-PELT	 &5.5   &0.3    &&9.0&0.3     &&11.5&0.3  \\   	
         &35         &	SMUCE	 &66.0  &1.8         &&71.0&2.3          &&77.5&2.8  \\
         &      	 &	SS-PELT	 &5.0  &0.7    &&11.5&0.7       &&19.5&0.7  \\
\bottomrule
\end{tabular}}
\caption{{{{ Performance study of SMUCE and our method SS-PELT, with $n=1000$ and SNR$=1.2$. Here the estimates $\widehat{K}_{\rm SMUCE}$ and $\widehat{K}_{\rm CV}$
for SMUCE and CV are denoted by $\widehat{K}$. For SMUCE, $P_+$ (in \%) is the empirical version of $\mathbb{P}(\widehat{K}_{\rm SMUCE}>K_n)$.
}}}}\label{tab:SMUCE}
\end{table}
\begin{table*}[t]      \tabcolsep 1.5pt
\centering
\linespread{1.3}
{\small \centering\begin{tabular}{cllcccccccccccccccccccccccc}
\toprule
&&& \multicolumn{2}{c}{$\alpha=5\%$} && \multicolumn{2}{c}{$\alpha=10\%$}&& \multicolumn{2}{c}{$\alpha=20\%$}\\
\cline{4-5}\cline{7-8} \cline{10-11}
 SNR&$K_n$&Method & $P_+$&$U$&&$P_+$&$U$&& $P_+$&$U$&&$\widehat{K}_{\rm CV}-K_n$\\
\hline	
    0.9&15       &	SS-WBS	  &3.0&0.9(1.6)    &&6.0&0.7(1.6)     &&7.0&0.7(1.5)  &&0.6(1.4) \\
   &         &	SS-PELT	  &0.0&0.9(1.3)    &&2.5&0.8(1.3)     &&2.0&0.4(0.9)  &&0.4(0.9) \\
   &         &	RC-PELT	  &0.0&0.5(1.0)    &&0.0&0.4(0.6)     &&0.0&0.3(0.6)  &&0.3(0.9) \\
   &25       &	SS-WBS	  &4.5&3.2(3.7)    &&10.5&2.5(3.2)    &&15.5&2.1(3.2) &&1.5(2.5)\\  		 	  		
   &         &	SS-PELT   &1.0&3.6(3.9)    &&2.5&2.3(2.4)     &&7.0&1.7(2.6)  &&0.6(1.2)\\
   &      	 &	RC-PELT	  &0.0&2.0(2.1)    &&0.0&1.2(2.0)     &&0.0&0.8(1.6)  &&0.5(1.1) \\
\\
1.2&15        &	SS-WBS	 &0.3  &0.7(1.4) &&3.0&0.6(1.2)       &&5.0&0.5(1.2)  &&0.6(1.4) \\
   &         &	SS-PELT	 &0.7  &0.4(1.0) &&1.5&0.4(0.8)       &&2.3&0.4(1.1)  &&0.3(1.0) \\
   &         &	RC-PELT	 &0.0  &0.4(0.8) &&0.0&0.3(0.8)      &&0.0&0.3(0.7)  &&0.4(1.0) \\
  &25 	     &	SS-WBS	 &4.5  &1.3(2.3) &&4.0&1.0(1.9)       &&7.0&0.9(1.8)  &&0.8(1.7) \\  	
   &         &	SS-PELT	 &0.0  &0.9(1.5) &&1.0&0.7(1.3)       &&3.0&0.6(1.3)  &&0.4(1.0) \\
   &      	 &	RC-PELT	 &0.0  &0.6(1.1) &&0.0&0.5(0.9)      &&0.0&0.4(0.8)
   &&0.4(0.9)\\
\bottomrule
\end{tabular}}
\caption{{{{ Performance study of the average value of $U$, $\widehat{K}_{\rm CV}-K_n$ and the empirical probability
$P_+$ (in \%), with $n=1000$ under Gaussian distribution. Standard deviations are listed in the brackets.}}}} \label{tab:M2}
\end{table*}
\begin{figure*}[!htb]\centering
\includegraphics[height=6cm,width=6cm]{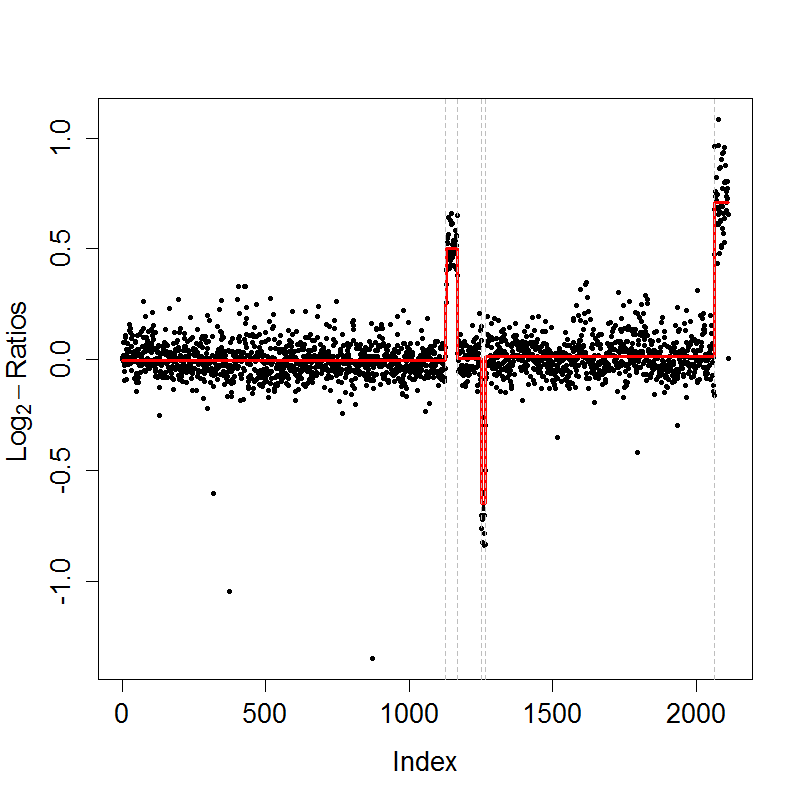}\includegraphics[height=6cm,width=6cm]{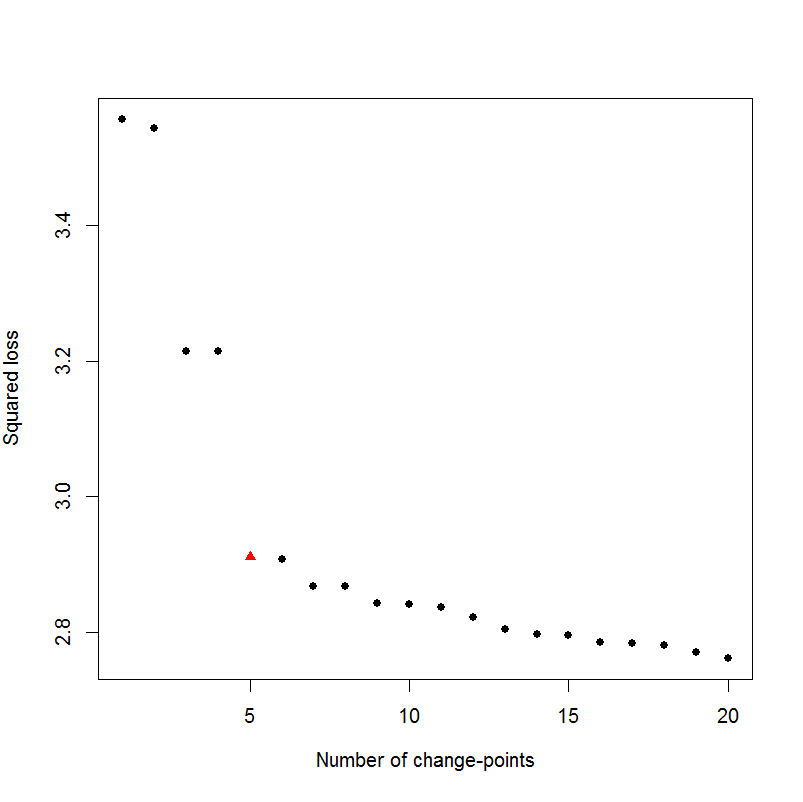}
\caption {\small\it Slope heuristics (right panel) of cell line GM05296 with the red solid line denoting the sample mean in each segmentation estimated by CV-WBS (left panel).}\label{example}
\end{figure*}
    Table \ref{tab:SMUCE} presents a comparative analysis between our proposed method and the benchmark approach SMUCE under normal and \(t(5)\) error term distributions. Our method consistently maintains \(P_+\) within the specified \(\alpha\) for both distributions. However, the benchmark, while controlling \(P_+\) for normal errors, exhibits instances of \(\widehat{K} - K_n\) being smaller than 0, indicating slight underestimation. Notably, for error terms with a \(t(5)\) distribution, the benchmark results in significantly higher \(P_+\) than the nominal level.

    Table \ref{tab:M2} assesses the performance of our methods combined with different change-point detection algorithms across various settings. All methods demonstrate the capability to control \(P_+\) at different nominal levels \(\alpha\).

\subsection{Real data}
Change-point detection is useful for detecting changes in a wide range of applications, such as bioscience, image analysis, and so on.
One area where change-point detection is commonly applied is in DNA copy number analysis.
For example, in array Comparative Genomic Hybridization (array CGH) data analysis, change-point detection is used to identify segments of the genome where there is a change in the copy number of DNA. The copy number refers to the number of copies of a specific DNA segment in a cell. By detecting these changes, researchers can identify regions of the genome that may be associated with genetic diseases or abnormalities.

Consider the array comparative genomic hybridization (CGH) data from the coriell dataset available in the R package \texttt{DNAcopy} in \citet{DNAcopy}.
Here the goal is to quantify the changes in terms of DNA copy number. We focus on the chromosomal aberration locations detection
and consider cell line GM05296 for illustration, which can be treated as a univariate mean shift detection problem (see the left panel in Figure \ref{example}).
There are totally $2112$ observations after removing na values.

Here we refer to \citet{lavielle2005using} and \citet{haynes2017computationally}, which suggest a method for determining the optimal number of change-points in a dataset by looking at how the cost changes as the number of change-points increases. The ``elbow" (red triangular point) in the right panel of Figure \ref{example} is used to locate the optimal number of change-points. We note that the cost should decrease more when detecting true changes, and in this case, the true number of change-points is determined to be $5$.

We investigate the performance of \(\widehat{K} - K_n\) for both our proposed method (CV-WBS) and the benchmark approach SMUCE. The result of CV-WBS is $1$. The benchmark yields a result of $18$ under a target level $\alpha$ of $0.05$, $0.1$ and $0.2$. Though both methods consistently yield nonnegative values for \(\widehat{K} - K_n\), CV-WBS provides significantly smaller values compared to the benchmark, indicating slight overestimation.

\section{Discussion}
In this study, we have presented a novel approach for quantifying uncertainty of cross-validation criterion in learning multiple change-points. Through the utilization of the model comparison framework and advanced high-dimensional inferential tools, we have established a robust methodology that allows us to control the probability of overestimation within a specified level. Our findings contribute to a better understanding of the reliability and robustness of the cross-validation criterion in change-point detection. Experimental results demonstrate the superiority of our proposed approach in providing accurate uncertainty quantification for overestimation compared to existing methods in finite samples.

One limitation of our method is its assumption of parametric and low-dimensional change-point models. Adapting the method to high-dimensional or nonparametric models and exploring its theoretical properties would be a valuable avenue for future research. Additionally, the assumption of independent observations may not be suitable for handling dependent data. Developing customized cross-validation frameworks and inferential tools for dependent data is an important direction to consider. By addressing these limitations and extending the method to encompass high-dimensional, non-parametric, and possibly correlated models, we can enhance its practical applicability and gain deeper insights into uncertainty quantification for cross-validation criterion in change-point detection in diverse real-world scenarios.

\section*{Acknowledgments}

The authors would like to express their sincere appreciation to the anonymous reviewers for their valuable comments and constructive feedback. Zou was supported by the National Key R\&D Program of China (Grant Nos. 2022YFA1003703, 2022YFA1003800) and the National Natural Science Foundation of China (Grant Nos. 11925106, 12231011, 11931001, 11971247). Chen was supported by the National Natural Science Foundation of China (Grant No. 12201257). Wang was supported by the National Natural Science Foundation of China (Grant No. 32030063) and the Natural Science Foundation of Shanghai (Grant No. 23ZR1419400).
\bigskip

\bibliography{aaai24}

\appendix
\newpage
\section{General results}

We first present theoretical results for the general model proposed in Section~\ref{subsec:ss-procedure}. Theorem~\ref{thmconsis} for the mean change model can be derived from Theorem~\ref{thm1}.

\begin{thm}\label{thm1}
Suppose Assumption \ref{assu:seup} holds. If $\max_{s> r}\frac{\Delta_{r,s}}{\sigma_{r,s}}\leq a_n(n\log n)^{-1/2}$, for some $a_n=o(1)$, then $\mathbb{P}( \mathcal{D}>c_{r, \alpha})\leq \alpha+o(1)$.
\end{thm}
If $H_{0,K_n}$ is accepted, we have $K_n\geq  \widehat{K}_{\rm min}$. Hence, by Theorem \ref{thm1}, the following corollary holds:

\begin{coro}\label{coro:consite} Suppose Assumption \ref{assu:seup} holds. If $\max_{s> K_n}\frac{\Delta_{K_n,s}}{\sigma_{K_n,s}}\leq a_n(n\log n)^{-1/2}$, for some $a_n=o(1)$, then
\begin{equation*}
    \begin{aligned}
    \mathbb{P}(  \widehat{K}_{\rm cv}-K_n>  U)\leq \alpha+o(1).
    \end{aligned}
\end{equation*}
\end{coro}

\section{Auxiliary lemmas}

In this section, we present some auxiliary lemmas. In the following content, $C$, $C'$ denote  positive constants which may vary across different scenarios.

\begin{lemma}\label{lemma:bernstein}[Bernstein's inequality]
%%%%%%%%%%%%%%%%%%%%%%%%%%%%%%%%%%%%%%%%%%%%%%%%%%%%%%%%%%%%
Let $X_1,\ldots,X_n$ be independent centered random variables a.s. bounded by $A<\infty$ in absolute value. Let $\sigma^2=n^{-1}\sum_{i=1}^n\E(X_i^2)$. Then for all $x>0$,
\begin{eqnarray*}
  \mathbb{P}\bigg(\sum_{i=1}^n X_i\geq x\bigg)\leq\exp\Big(-\frac{x^2}{2n\sigma^2+2Ax/3}\Big).
\end{eqnarray*}
\end{lemma}

\begin{lemma}\label{lemmasplit}
Suppose Assumption \ref{assu:seup} (i) holds. For any $0<\nu_1<1$ and $0<\varsigma <\vartheta-2$, the following facts hold:
\begin{align}
&\mathbb{P}\left(\max_{s> r}\left|\frac{  \widehat{\Delta}_{r,s}-\Delta_{r,s}}{\sigma_{r,s}}\right|\geq Cn^{-1/2}\sqrt{\log p_n}\right)\leq C'n^{1-\frac{\vartheta}{\vartheta-\varsigma}},\label{spliteq1}\\
&\mathbb{P}\bigg\{\max_{s> r}\left|\frac{  \widehat{\sigma}_{r,s}}{\sigma_{r,s}}-1\right|\geq C(n^{(\nu_1-1)/2}B_{n}^2\log {p_n}\nonumber\\
&~~~~~~~~~~~~~~~~~~~~~~~~~~~+n^{-3/2}B_{n}^2\log^2{p_n})\bigg\}\leq C'n^{-\nu_1},\label{spliteq2}\\
&\mathbb{P}\Big\{\max_{s> r, s> r'}|  \widehat{{\bm\Gamma}}_{r;s,s'}-\bm\Gamma_{r;s,s'}|\geq  C(n^{(\nu_1-1)/2}B_{n}^2\log {p_n}\nonumber\\
&~~~~~~~~~~~~~~~~~~~~~~~~~~~+n^{-3/2}B_{n}^2\log^2{p_n})\Big\}\leq C'n^{-\nu_1},\label{spliteq3}
\end{align}
where $\bm\Gamma_{s,s'}$ is the correlation coefficient between $\delta^{(i)}_{r,s}$ and $\delta^{(i)}_{r,s'}$.
\end{lemma}
{\bf{Proof of Lemma \ref{lemmasplit}}.} For simplicity, we fix $r$ and drop $r$ in the notation.  We first  prove (\ref{spliteq1}).
Denote $M_n=n^{1/(\vartheta-\varsigma)}$ for some $0<\varsigma<\vartheta$, and $\phi^{(i)}_s=\{\delta^{(i)}_{r,s}-\mathbb{E}(\delta^{(i)}_{r,s}\mid{\bm\xi}^{\rm tr})\}/\sigma_{r,s}$. Note that
\begin{align*}
\phi^{(i)}_s&=\phi^{(i)}_s{\bm 1}(|\phi^{(i)}_s|\leq M_n)-\mathbb{E}\{\phi^{(i)}_s{\bm 1}(|\phi^{(i)}_s|\leq M_n)\}\\
&~~~+\phi^{(i)}_s{\bm 1}(|\phi^{(i)}_s|> M_n)-\mathbb{E}\{\phi^{(i)}_s{\bm 1}(|\phi^{(i)}_s|> M_n)\}\\
   &=:\phi^{(i1)}_s+\phi^{(i2)}_s.
\end{align*}
We have
\begin{align*}
&~~~\mathbb{P}\left(\max_{s} n^{-1/2}\left|\sum_{i=1}^{n}\phi^{(i)}_s\right|>x\right)\\
&\leq\mathbb{P}\left(\max_{s} n^{-1/2}\left|\sum_{i=1}^{n}{\phi^{(i1)}_s}\right|>x/2\right)\\
&~~~+\mathbb{P}\left(\max_{s} n^{-1/2}\left|\sum_{i=1}^{n}{\phi^{(i2)}_s}\right|>x/2\right)\\
&=:P_1+P_2.
\end{align*}

Set $x=\sqrt{C\log n}$ for a sufficiently large $C$. We derive the upper bound for $P_1$. By Lemma \ref{lemma:bernstein}, we have
\begin{align*}
P_1&=p_n\mathbb{P}\left(n^{-1/2}\left|\sum_{i=1}^{n}{\phi^{(i1)}_s}\right|>x/2\right)\\
&\leq p_n\exp\left\{-\frac{nx^2}{C_1n+C_2M_nn^{1/2}x}\right\}=o(n^{1-\frac{\vartheta}{\vartheta-\varsigma}})
\end{align*}
by $\varsigma<\vartheta-2$, where $C_1,C_2$ are some positive constants.

On the other hand, note that
\begin{align*}
P_2\leq& \mathbb{P}\bigg[\max_{s}n^{-1/2}\sum_{i=1}^{n}|{\phi^{(i)}_s}|{\bm 1}(|{\phi^{(i)}_s}|> M_n)\\
&+\max_{s}n^{1/2}\mathbb{E}\{|{\phi^{(i)}_s}|{\bm 1}(|{\phi^{(i)}_s}|> M_n)\}>x/2\bigg].
\end{align*}
By the Cauchy-Schwarz inequality and Markov's inequality, we observe that
\begin{align*}
\mathbb{E}^2\{|{\phi^{(i)}_s}|{\bm 1}(|{\phi^{(i)}_s}|> M_n)\}&\leq \mathbb{E}({\phi^{(i)}_s})^2\mathbb{P}(|{\phi^{(i)}_s}|> M_n)\\
&\leq M_n^{-\vartheta}\mathbb{E}({\phi^{(i)}_s})^2\mathbb{E}(|{\phi^{(i)}_s}|^{\vartheta}),
\end{align*}
which yields
\begin{align*}
\max_{s}n^{1/2}\mathbb{E}\{|{\phi^{(i)}_s}|{\bm 1}(|{\phi^{(i)}_s}|> M_n)\}=o(1).
\end{align*}
Thus, we have
\begin{align*}
P_2\leq& \mathbb{P}\bigg\{\max_{s}n^{-1/2}\sum_{i=1}^{n}|{\phi^{(i)}_s}|{\bm 1}(|{\phi^{(i)}_s}|> M_n)>x/4\bigg\}\\
\leq&nM_n^{-\vartheta}\mathbb{E}|\phi^{(i)}_s|^{\vartheta}=O(n^{1-\frac{\vartheta}{\vartheta-\varsigma}}).
\end{align*}
Hence, (\ref{spliteq1}) is proved. 

See Lemma D.5 in \citet{chernozhukov2019inference} for the proof of (\ref{spliteq2}).

Next, we derive (\ref{spliteq3}).
By the Cauchy-Schwarz inequality, we have
\[\mathbb{E}\left(\left|\frac{(\delta_s^{(i)}-\Delta_s)(\delta_{s'}^{(i)}-\Delta_{s'})}{\sigma_s\sigma_{s'}}\right|^2\right)\leq B_{n}^4.\]

Also,
\begin{align*}
    &~~~\mathbb{E}\left(\max_{i;s,s'}\left|\frac{(\delta_s^{(i)}-\Delta_s)(\delta_{s'}^{(i)}-\Delta_{s'})}{\sigma_s\sigma_{s'}}\right|^2\right)\\&\leq \mathbb{E}\left(\max_{i;s}\left|\frac{(\delta_s^{(i)}-\Delta_s)}{\sigma_s}\right|^2\max_{i;s'}\left|\frac{(\delta_{s'}^{(i)}-\Delta_{s'})}{\sigma_{s'}}\right|^2\right)\\&\leq nB_{n}^4.
\end{align*}
Since
\begin{align*}
      &~~~\widehat{\bm\Gamma}_{s,s'}-\bm\Gamma_{s,s'}\\&=\frac{1}{n_{\rm te}}\sum_{i\in\mathcal{I}_{\rm te}}\left\{\frac{(\delta_s^{(i)}-\Delta_s)(\delta_{s'}^{(i)}-\Delta_{s'})}{\sigma_s\sigma_{s'}}-\bm\Gamma_{s,s'}\right\}\frac{\sigma_s\sigma_{s'}}{  \widehat{\sigma}_s  \widehat{\sigma}_{s'}}\\
      &~~~+\bm\Gamma_{s,s'}\left(\frac{\sigma_s\sigma_{s'}}{  \widehat{\sigma}_s  \widehat{\sigma}_{s'}}-1\right)-\frac{(  \widehat{\Delta}_s-\Delta_s)(  \widehat{\Delta}_{s'}-\Delta_{s'})}{\sigma_s\sigma_{s'}}\frac{\sigma_s\sigma_{s'}}{  \widehat{\sigma}_s  \widehat{\sigma}_{s'}},
\end{align*}
by Lemma D.2 and Lemma D.3 in \citet{chernozhukov2019inference}, similar to the proof of (\ref{spliteq2}), (\ref{spliteq3}) follows.

\section{Proof of Theorem~\ref{thm1}}

In this section, 
we fix $r$ and drop $r$ in the notation. For any positive semidefinite matrix $\bf M$, let $Z_{\bf M}$ be an $N(0,{\bf M})$ random vector. For $\alpha\in(0,1)$, let $z(\alpha,{\bf M})$ be the upper $\alpha$ quantile of the maximum of $Z_{\bf M}$. 

Define the event
\begin{align*}
\mathcal{E}&=\Big\{\max_{s}|\frac{  \widehat{\Delta}_s-\Delta_s}{\sigma_s}|\leq Cn^{-1/2}\sqrt{\log p_n}\Big\}\\
&\cap\Big\{\max_{s}|\frac{  \widehat{\sigma}_s}{\sigma_s}-1|\leq C(n^{-(1-\nu_1)/2}B_{n}^2\log p_n\\
&~~~~~~~~~+n^{-3/2}B_{n}^2\log^2p_n)\Big\}\\
&\cap\Big\{\max_{s,s'}|  \widehat{\bm\Gamma}_{s,s'}-\bm\Gamma_{s,s'}|\leq C(n^{-(1-\nu_1)/2}B_{n}^2\log p_n\\
&~~~~~~~~~+n^{-3/2}B_{n}^2\log^2p_n)\Big\}.
\end{align*}
 Denote $\psi_n=a_n\{n\log n\}^{-1/2}$. We have
% under $H_{0,r}$, as $B\rightarrow\infty$,
\begin{align*}
    &~~~\mathbb{P}(\mathcal{D}>c_{\alpha})\nonumber\\ &=\mathbb{P}\bigg\{\sqrt{n_{\rm te}}\max_{s}\frac{  \widehat{\Delta}_s}{  \widehat{\sigma}_s}\geq z(\alpha,  \widehat{\bm\Gamma})\bigg\}\nonumber\\
    &\leq \mathbb{P}\bigg\{\sqrt{n_{\rm te}}\max_{s}\frac{  \widehat{\Delta}_s-\Delta_s}{\sigma_s}\left(\frac{\sigma_s}{  \widehat{\sigma}_s}-1\right)+\sqrt{n_{\rm te}}\max_{s}\frac{  \widehat{\Delta}_s-\Delta_s}{\sigma_s}\\
    &~~~~~~~~~\geq z(\alpha,  \widehat{\bm\Gamma})-c\sqrt{n}\psi_n,\mathcal{E}\bigg\}+\mathbb{P}(\mathcal{E}^{c})\nonumber\\
    &\leq \mathbb{P}\bigg\{\sqrt{n_{\rm te}}\max_{s}\frac{  \widehat{\Delta}_s-\Delta_s}{\sigma_s}\geq z(\alpha,  \widehat{\bm\Gamma})-c\sqrt{n}\psi_n\\
    &~~~~~~~~~-Cn^{(\nu_1-1)/2}B_{n}^2\log^{3/2}p_n-Cn^{-3/2}B_n^2\log^{5/2}p_n,\mathcal{E}\bigg\}\\
    &+\mathbb{P}(\mathcal{E}^{c})\nonumber,
\end{align*}
for some constant $c>0$. 
Denote $\varrho=c\max\{n^{-(1-\nu_1)/6}B_{n}^{2/3}\log p_n, n^{-1/2}B_n^{2/3}\log^{4/3}p_n\}$ and $\nu_2=\min(\frac{\vartheta}{\vartheta-\varsigma}-1, \nu_1)$. By Theorem 2 in \citet{chernozhukov2015comparison}, we have
\[z(\alpha,  \widehat{\bm\Gamma})\geq z(\alpha+\varrho,\bm\Gamma).\]
Hence,
\begin{align*}
    &~~~\mathbb{P}(\mathcal{D}>c_{\alpha})\nonumber\\
    &\leq \mathbb{P}\bigg\{\sqrt{n_{\rm te}}\max_{s}\frac{  \widehat{\Delta}_s-\Delta_s}{\sigma_s}\geq z(\alpha+\varrho,\bm\Gamma)-c\sqrt{n}\psi_n\\
    &~~~~~~~~~-Cn^{(\nu_1-1)/2}B_{n}^2\log^{3/2}p_n-Cn^{-3/2}B_n^2\log^{5/2}p_n,\mathcal{E}\bigg\}\nonumber\\
    &~~~+\mathbb{P}(\mathcal{E}^{c})\nonumber\\
    &\leq \mathbb{P}\Big\{\max Z_{\bm\Gamma}\geq z(\alpha+\varrho,\bm\Gamma)-c\sqrt{n}\psi_n\\
    &~~~~~~~~~-Cn^{(\nu_1-1)/2}B_{n}^2\log^{3/2}p_n-Cn^{-3/2}B_n^2\log^{5/2}p_n\Big\}\\
    &~~~+C'n^{-c_1}+C'n^{-\nu_2},
\end{align*}
by combining Corollary 2.1 in \citet{chernozhukov2013gaussian} for some $c_1>0$. By (\ref{assu1}) and using anti-concentration of maxima of mean vectors (cf. Theorem 3 in \citet{chernozhukov2015comparison}), it follows that
\begin{align*}
    &~~~\mathbb{P}(\mathcal{D}>c_{\alpha})\nonumber\\
    &\leq \alpha+\varrho\\
    &~~~+(c\sqrt{n}\psi_n+Cn^{(\nu_1-1)/2}B_{n}^2\log^{3/2}p_n+Cn^{-3/2}B_n^2\log^{5/2}p_n)\\
    &~~~\times(1+\sqrt{2\log p_n})+C'n^{-c_1}+C'n^{-\nu_2}\\
    &=\alpha+o(1).
\end{align*}

\section{Proof of Theorem \ref{thmconsis}}
We first introduce some notations.
Let $\{{\bf x}_1,\ldots,{\bf x}_n\}$ and $\{{\bf y}_1,\ldots,{\bf y}_n\}$ be two sets of $d$-variate vectors. Let $\mathcal{T}_r=(\tau_1,\ldots,\tau_r)$ be a set of $r$ points such that $0<\tau_1<\cdots<\tau_r<n$. 
We denote
\begin{align*}
    \mathcal{S}_{{\bf x}{\bf y}}(\mathcal{T}_r)=&\sum_{k=0}^r\sum_{i=\tau_k+1}^{\tau_{k+1}}({\bf x}_i-\bar{\bf x}_{\tau_k,\tau_{k+1}})^{\top}({\bf y}_i-\bar{\bf y}_{\tau_k,\tau_{k+1}}),
\end{align*}
where $\tau_0=0$ and $\tau_{r+1}=n$. Define  $\mathcal{S}_{{\bf x}}^2=\mathcal{S}_{{\bf x}{\bf x}}$.

The main goal of our proof is to show that $\mathbb{P}(  \widehat{K}_{\rm min}\geq K_n)$ is small enough. The event $\{\widehat{K}_{\rm min}\geq K_n\}$ means that for any $r< K_n$, the hypotheses  $H_{0, r}$'s are all rejected. If $H_{0, K_n}$ is accepted, then $K_n\geq   \widehat{K}_{\rm min}$. By Theorem \ref{thm1},  we can prove  that $H_{0, K_n}$ is accepted, and 
\begin{equation*}
    \begin{aligned}\label{equ:apdx:alpha}
    \mathbb{P}(  \widehat{K}_{\rm min}\geq K_n)\leq \mathbb{P}(\mathcal{D}>c_{K_n, \alpha})\leq \alpha+o(1).
    \end{aligned}
\end{equation*}
Hence, we need to check the condition in Theorem \ref{thm1}.  It suffices to prove that 
\begin{equation}
    \begin{aligned}\label{equ:apdx:overfit}
         \mathbb{P}(\max_{s>K_n}\Delta_{K_n,s}/\sigma_{K_n,s}\leq0)=1.
    \end{aligned}
\end{equation}

We first compute $\Delta_{{K_n},s}$. Since random vectors $\bm\varepsilon_1,\ldots,\bm\varepsilon_n$ are independent with mean zero, we have
\begin{align*}
    \mathbb{E}\{\mathcal{S}_{\bm\varepsilon^{\rm tr}\bm\varepsilon^{\rm te}}(  \widehat{\mathcal{T}}_s)\mid  \widehat{\mathcal{U}}_{s}^{\rm tr},  \widehat{\mathcal{U}}_{{K_n}}^{\rm tr}\}=0.
\end{align*}
Due to the data splitting scheme, we have $\bm\mu_i^{\rm tr}=\bm\mu_i^{\rm te}$ for each $i\in\mathcal{I}_{\rm te}$. Hence, 
we can decompose $n_{\rm te}\Delta_{K_n, s}$ as follows, 
\begin{equation*}
    \begin{aligned}\label{equ:decompose-delta}
    &~~~n_{\rm te}\Delta_{K_n,s}\\
  &=\mathcal{C}_{\bm\xi^{\rm tr}}^2(\widehat{\mathcal{T}}_{K_n})-\mathcal{C}_{\bm\xi^{\rm tr}}^2(\widehat{\mathcal{T}}_s) - 2\{\mathcal{C}_{\bm\mu^{\rm tr}\bm\xi^{\rm tr}}(\widehat{\mathcal{T}}_{K_n})-\mathcal{C}_{\bm\mu^{\rm tr}\bm\xi^{\rm tr}}(\widehat{\mathcal{T}}_{s})\}.
    \end{aligned}
    \end{equation*}
By condition (\ref{equ:variation}), (\ref{equ:apdx:overfit}) holds. Then, 
\begin{align*}
     \mathbb{P}(\widehat{K}_{\rm cv}-K_n>U)\leq \alpha+o(1).
\end{align*}

\section{Proof of Proposition~\ref{prop:underestimation}}

(i) Note that $\widehat{K}_{\rm CV}=\argmin_{1\leq s\leq p_n}n_{\rm te}\widehat{\Delta}_{K_n, s}$, we always have $\mathcal{D}\leq 0$. If $\alpha<0.5$, as $B\to\infty$, the upper $\alpha$ quantile of the maximum of a zero-mean Gaussian random vector $(\mathcal{D}_b)$ is positive. Hence, we have $\lim_{B\to\infty}\mathbb{P}(\widehat{K}_{\rm CV}\geq \widehat{K}_{\rm min})=1$ when $\alpha<0.5$.

(ii)  
To prove $\lim_{n\to\infty}\mathbb{P}(\widehat{K}_{\rm CV}\geq K_n)=1$, it suffices to show that the validation error for the change-points model $\widehat{\mathcal{U}}_s^{\rm tr}$ is larger than that for the model $\widehat{\mathcal{U}}_{K_n}^{\rm tr}$ for $s<K_n$.

We can decompose $n_{\rm te}\widehat{\Delta}_{K_n, s}$ as follows, 
\begin{align*}
    n_{\rm te}\widehat{\Delta}_{K_n,s}&=\{\mathcal{S}_{\bm\xi^{\rm te}}^2(  \widehat{\mathcal{T}}_{K_n})-\mathcal{S}_{\bm\xi^{\rm te}}^2(  \widehat{\mathcal{T}}_{s})\}\\
    &~~~-\{\mathcal{S}^2_{\bm\varepsilon^{\rm te}}(  \widehat{\mathcal{T}}_{K_n})-\mathcal{S}^2_{\bm\varepsilon^{\rm te}}(  \widehat{\mathcal{T}}_s)\}\\
  &~~~+2\{\mathcal{S}_{\bm\varepsilon^{\rm tr}\bm\varepsilon^{\rm te}}(  \widehat{\mathcal{T}}_{K_n})-\mathcal{S}_{\bm\varepsilon^{\rm tr}\bm\varepsilon^{\rm te}}(  \widehat{\mathcal{T}}_s)\}\\
  &~~~-\{\mathcal{S}^2_{\bm\varepsilon^{\rm tr}}(  \widehat{\mathcal{T}}_{K_n})-\mathcal{S}^2_{\bm\varepsilon^{\rm tr}}(  \widehat{\mathcal{T}}_{s})\}.
\end{align*}
By Fact (A), Fact (B) and Lemma 4 in \citet{zou2020consistent}, 
\begin{align*}
    \mathcal{S}_{\bm\xi^{\rm te}}^2(  \widehat{\mathcal{T}}_{s})-\mathcal{S}_{\bm\xi^{\rm  te}}^2(  \widehat{\mathcal{T}}_{{K_n}})\geq &\frac{\underline{\lambda}_n}{8}\underline{\gamma}_n\{1+o_p(1)\},\\
\mathcal{S}^2_{\bm\varepsilon^{\rm te}}(  \widehat{\mathcal{T}}_s)-\mathcal{S}^2_{\bm\varepsilon^{\rm te}}(  \widehat{\mathcal{T}}_{{K_n}})=&O_{p}({K_n}\overline{\sigma}),\\
    \mathcal{S}_{\bm\varepsilon^{\rm tr}\bm\varepsilon^{\rm te}}(  \widehat{\mathcal{T}}_{K_n})-\mathcal{S}_{\bm\varepsilon^{\rm tr}\bm\varepsilon^{\rm te}}(  \widehat{\mathcal{T}}_s)=&O_p(K_n\overline{\sigma}\overline{\lambda}_n^{2/m}),\\
\mathcal{S}^2_{\bm\varepsilon^{\rm tr}}(  \widehat{\mathcal{T}}_s)-\mathcal{S}^2_{\bm\varepsilon^{\rm tr}}(  \widehat{\mathcal{T}}_{{K_n}})=&O_{p}({K_n}\overline{\sigma}\overline{\lambda}_n^{2/m}).
\end{align*}
By (\ref{equ:underfit}), 
we have $\lim_{n\rightarrow\infty}\mathbb{P}(n_{\rm te}\widehat{\Delta}_{K_n, s}<0)=1$. Hence, underestimation cannot happen. The conclusion follows.

\end{document}